\documentstyle[epsf]{article}
\def\slashchar#1{\setbox0=\hbox{$#1$}           
   \dimen0=\wd0                                 
   \setbox1=\hbox{/} \dimen1=\wd1               
   \ifdim\dimen0>\dimen1                        
      \rlap{\hbox to \dimen0{\hfil/\hfil}}      
      #1                                        
   \else                                        
      \rlap{\hbox to \dimen1{\hfil$#1$\hfil}}   
      /                                         
   \fi}                                        %

\begin{document}
\title{
{\small
\vspace{-.5in}
\begin{flushright}
ANL--HEP--PR--97--27
\end{flushright}
}
\vspace{.5in}
Physics Backgrounds to Supersymmetric Signals with Two Photons
and Missing Mass at LEP}
\author{S. Mrenna\cite{mrenna_email}\\
High Energy Physics Division\\
Argonne National Laboratory \\
Argonne, IL  60439}
\date{\today}
\maketitle
\begin{abstract}
\noindent
We calculate the event rates
for the Standard Model
production of two photons plus missing mass through
the process $e^+e^-\to\gamma\gamma\nu\bar{\nu}$, where
$\nu$ is any of the three neutrino flavors.  This process
can be a background to new physics signatures, such as
that expected from supersymmetry with a higgsino or
gravitino LSP.  The missing mass spectrum is presented 
for several sets of kinematic cuts at $\sqrt{s}$=161, 172,
184, and 194 GeV.  The effect of initial state photon
radiation is also studied, and is found to be significant,
approximately doubling the event rates.
\end{abstract}
%
%
%
\section{Introduction}
\indent

The general case for weak scale supersymmetry has been reviewed in
Ref.\cite{snowmass}.
Recently, some evidence has accumulated that is
consistent with a light supersymmetric particle spectrum
containing a lightest superpartner (LSP) that is the
lightest neutralino $\tilde N_1$ with a high higgsino
content\cite{us}.  
In such models, a photino--like neutralino $\tilde N_2$ can
decay $\tilde N_2\to \tilde N_1 \gamma$, where $\tilde N_1$
escapes the detector.  Direct pair production of $\tilde N_2$
or the indirect pair production from the decays of
sneutrino pairs, $\tilde \nu (\to \tilde N_2 \nu)$, 
can lead to a striking
signature $\gamma\gamma \slashchar{M}$, where $\slashchar{M}$
is missing mass.  Such signatures are also possible from neutralino
production and decay for a gravitino LSP $\tilde G$, e.g. $e^+e^-\to
\tilde N_1(\to \gamma\tilde G) \tilde N_1(\to\gamma\tilde G)$, in
models of gauge mediated supersymmetry breaking\cite{gmsb}.
See also the references \cite{more}.

In the Standard Model, similar final states can arise from
the process $e^+e^-\to\gamma\gamma\nu\bar{\nu}$.  When
$\nu=\nu_\mu,\nu_\tau$, the neutrinos are produced through
a $Z$ boson resonance.  Hence, the $\slashchar{M}$ distribution
has a characteristic shape peaked at $M_Z$.
However, when $\nu=\nu_e$, there are
additional contributions from virtual $W$ bosons.  As a result,
the $\nu_e$ process acquires a tail in the $\slashchar{M}$ 
distribution far from the $Z$ boson mass. 

The purpose of this report is to quantify the Standard Model rate
for $\gamma\gamma\slashchar{M}$ events, so that one can address the
significance of any such events with large $\slashchar{M}$.


\section{Calculation and Results without Initial State Radiation}
\label{sec2}
\indent

The cross section for
the process $e^+e^-\to\gamma\gamma\nu\bar{\nu}$ is calculated
using the helicity amplitude library
{\tt HELAS}\cite{helas} to evaluate amplitudes derived by the 
program {\tt Madgraph}\cite{madgraph}.  
The inputs to the program are $\alpha=.0078$, $\sin\theta_W=.230$,
$M_W=80$ GeV, $\Gamma_W=1.903$ GeV, $M_Z=91.17$ GeV,
and $\Gamma_Z=2.288$ GeV.  The calculation is done at a fixed order,
with two and only two hard photons in the final state.  The effect
of undetected initial state radiation is considered in Sec. 3.

Based on the analysis presented by the OPAL collaboration\cite{opal}, 
we require that both
photons satisfy $|\cos\theta|<.7$, where $\theta$ is the photon
angle with respect to the beam axis in the laboratory frame,
and $E^\gamma > E^\gamma_{\rm min}$, where $E^\gamma$ is the
photon energy.  We study the sensitivity of the event rate
and distributions to the choices $E^\gamma_{\rm min}$=1.75, 5, and 10
GeV and to another cut on $|\cos\theta|$.

Our results are summarized in Figures 1--4.
In Fig. 1, we present the missing mass distribution in pb per GeV bin
for a minimum photon energy of 1.75 GeV at
$\sqrt{s}$=161, 172, 184, and 194 GeV.  
The missing mass $\slashchar{M}$ is defined to be the invariant
mass of the four vector $(p_{e^+}+p_{e^-}-p_{\gamma_1}-p_{\gamma_2})^\mu$.
For $\slashchar{M}<80$ GeV, the cross section is negligible.
The integrated
cross section for $\slashchar{M}>100$ GeV is also shown in the figure.
For the machine energies considered, this number is almost a
constant, about .016 pb.  Figs. 2 and 3 show the same distributions for 
$E^\gamma_{\rm min}$=5 and 10 GeV respectively.  The integrated
cross section for $\slashchar{M}>100$ GeV is roughly .007 and .003
pb for the two cuts regardless of the beam energy.
Finally, Fig. 4
shows the correlation between the two photon energies
in pb per GeV$^2$ bin at $\sqrt{s}$=161 GeV.  There is a peak 
when both photons are soft, and a ridge for one soft and one hard
photon.  Away from this region, the distribution is fairly flat.

If, instead, we require $|\cos\theta|<.95$, the results are similar,
except that the event rate is higher.  For $E^\gamma_{\rm min}=$1.75, 5,
and 10 GeV, the integrated cross section for $\slashchar{M}>100$ GeV is
roughly .074, .033, and .016 pb regardless of the beam energy.
These numbers are in agreement with an independent
calculation\cite{sandro}.
Therefore, there is a substantial gain in reducing the Standard Model
contribution to $\gamma\gamma\slashchar{M}$ events by focussing on the
central region of the detector.

\section{Calculation and Results with Initial State Radiation}
\indent

In principle, the colliding $e^+$ and $e^-$ need not each have the beam 
energy, but may have radiated many undetected photons beforehand.
The effect of undetected initial state radiation is to reduce the
center of mass energy closer to $M_Z$ and to generate a component of
missing momentum down the beampipe.  The increasing partonic cross
section must be folded with the decreasing probability that the incoming
$e^-$ ($e^+$) carries a reduced fraction of the beam energy.  
In QED, the electron momentum distribution functions are 
entirely calculable, and
depend on the inputs $\alpha(\simeq 1/137)$, $m_e$, and the beam energy.
We use the next--to--leading--order exponentiated form as described in
Ref.\cite{kle89} and implemented in the {\tt PYTHIA} subroutine
{\tt PYPDEL}\cite{pythia}.

The effect of undetected initial state photon radiation 
is illustrated in Figs. 5--8.
In Fig. 5, we present the missing mass distribution in pb per GeV bin
for a minimum photon energy of 1.75 GeV at
$\sqrt{s}$=161, 172, 184, and 194 GeV.  
The integrated
cross section for $\slashchar{M}>100$ GeV 
at $\sqrt{s}=161$ GeV is .035 pb, or more than double the previous estimate
without initial state radiation.  This rate decreases slightly for
higher beam energies.
Figs. 6 and 7 show the same distributions for 
$E^\gamma_{\rm min}$=5 and 10 GeV respectively.  The integrated
cross section for $\slashchar{M}>100$ GeV is roughly .016 and .008
pb for the two cuts regardless of the beam energy, also about double
the previous estimate.
Finally, Fig. 8
shows the correlation between the two photon energies
in pb per GeV$^2$ bin at $\sqrt{s}$=161 GeV.  

If $|\cos\theta|<.95$, 
the integrated cross section for $\slashchar{M}>100$ GeV is
roughly .155, .076, and .035 pb at $\sqrt{s}=161$ GeV
for $E^\gamma_{\rm min}=$1.75, 5,
and 10 GeV, respectively.
As for the case of harder cuts, the rates are about double when including
initial state radiation.

Note that such calculation should not be trusted when one or
both of the photons are very soft.  In that case, multiple
soft photon emission will be important.
Likewise, event generators for a single hard photon and soft photons
should not be used to estimate the rate for two hard photons.
Finally, other potential Standard Model contributions to the
$\gamma\gamma\slashchar{M}$ final state are $e^+e^-\to\gamma\gamma
e^+e^-$, where both the electron and positron are undetected in the
beam pipe, or $e^+e^-\to\gamma\gamma X$, where $X$ is any number of 
undetected photons.   Such events can be rejected by demanding that 
the missing energy vector {\bf not} point down the beampipe and that
the photons are not back--to--back in the azimuthal plane.

\section{Conclusions}
\label{sec5}
\indent

We have calculated event rates and kinematic distributions
for the Standard Model
production of two photons plus missing mass through
the process $e^+e^-\to\gamma\gamma\nu\bar{\nu}$ at LEP
with beam energies
$\sqrt{s}$=161, 172, 184, and 194 GeV.  In particular, the
calculation includes important contributions from $W$ boson
exchange when $\nu=\nu_e$.  After applying a missing mass
$\slashchar{M}$ cut of 100 GeV and an angular acceptance cut
$|\cos\theta|<.7$, the integrated cross section
is fairly independent of the beam energy.  For a minimum photon
energy of 1.75, 5, and 10 GeV, the expected rate is .016, .007,
and .003 pb, respectively.  
These rates are approximately doubled after including the
effect of undetected initial state photon radiation.
The photon energies are highly
correlated when one or both are soft, but uncorrelated when
both are hard.  The kinematic region with a large $\slashchar{M}$
and two hard photons appears ideal to search for non--Standard
Model physics, such as supersymmetry with a higgsino--like or
gravitino LSP.

\section*{Acknowledgements}

We thank G.L.~Kane for encouragement in presenting these results,
and G.L.K. and G. Mahnon for comments.
This work was supported by DOE grant W--31--109--ENG--38.

%
%
%
\begin{figure}
\centering
\hspace*{0in}
\epsfxsize=5.0in
\epsffile{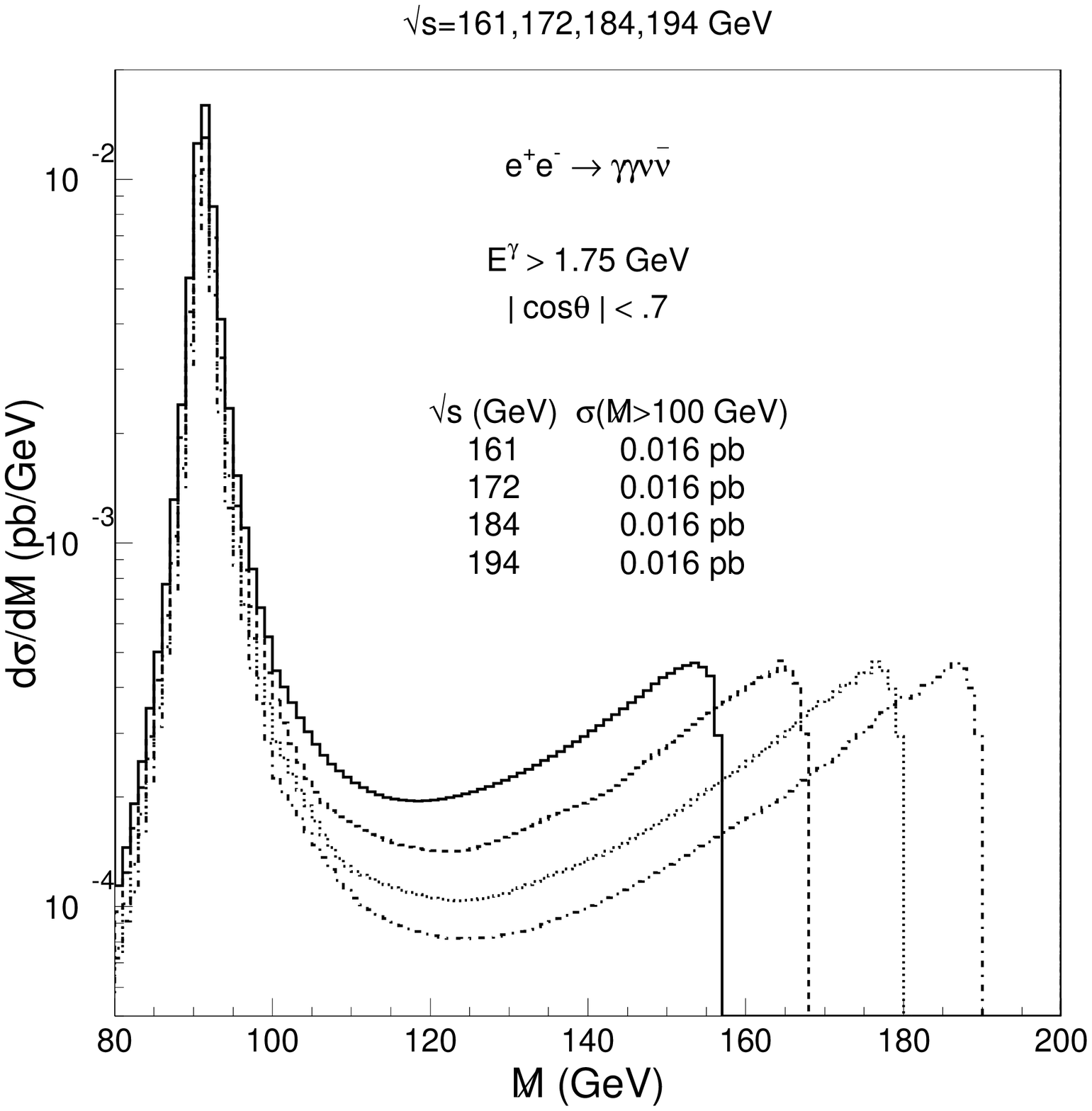}
\vspace*{0in}
\caption{The Missing Mass distribution in pb per GeV bin
for a minimum photon energy of 1.75 GeV and $|\cos\theta|<.7$ at
$\sqrt{s}$=161, 172, 184, and 194 GeV.  The integrated
cross section above 100 GeV is also shown.}
\label{fig_one}
\end{figure}
\begin{figure}
\centering
\hspace*{0in}
\epsfxsize=5.0in
\epsffile{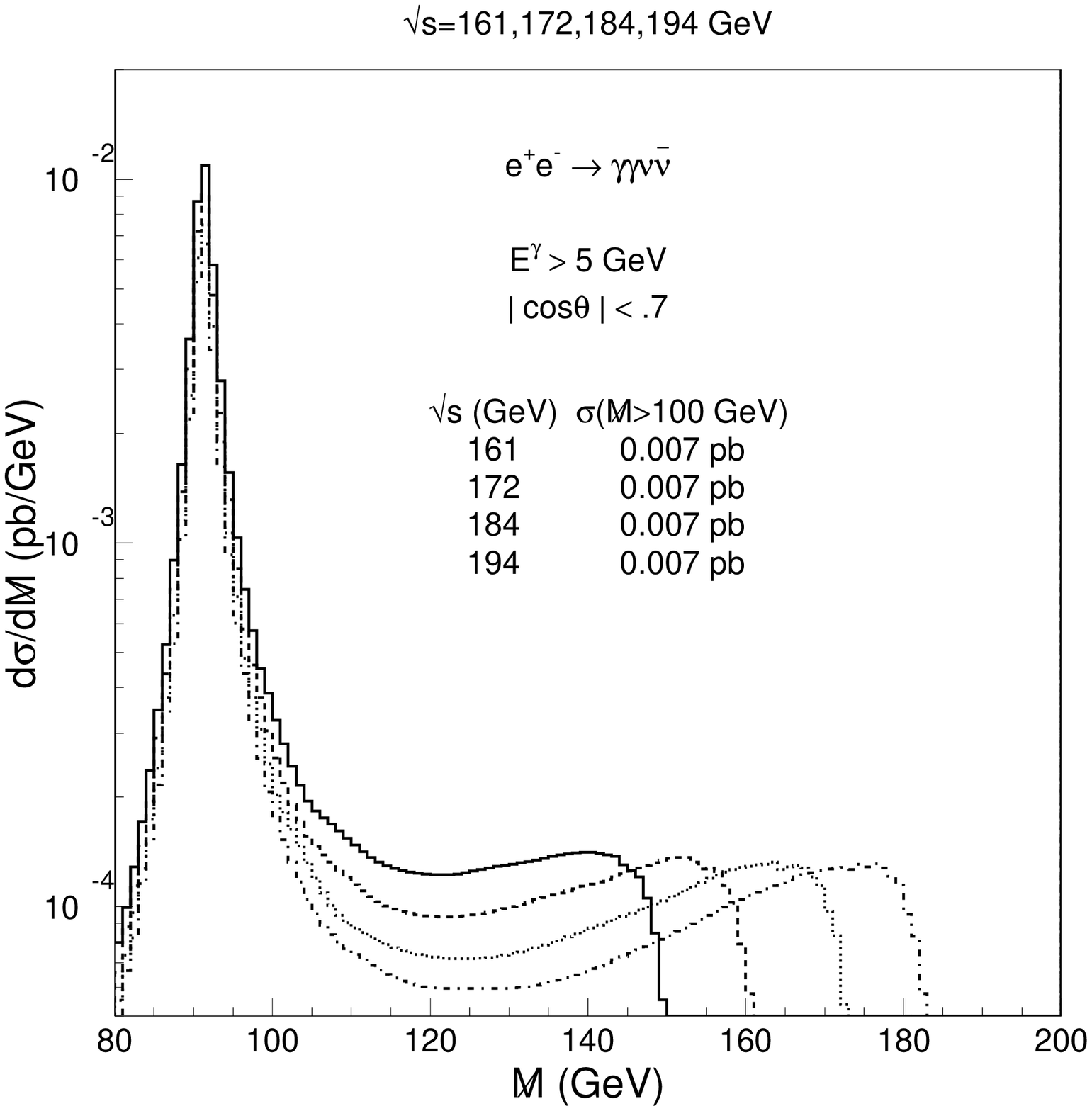}
\vspace*{0in}
\caption{The Missing Mass distribution in pb per GeV bin
for a minimum photon energy of 5 GeV and $|\cos\theta|<.7$ at
$\sqrt{s}$=161, 172, 184, and 194 GeV.  The integrated
cross section above 100 GeV is also shown.}
\label{fig_two}
\end{figure}
\begin{figure}
\centering
\hspace*{0in}
\epsfxsize=5.0in
\epsffile{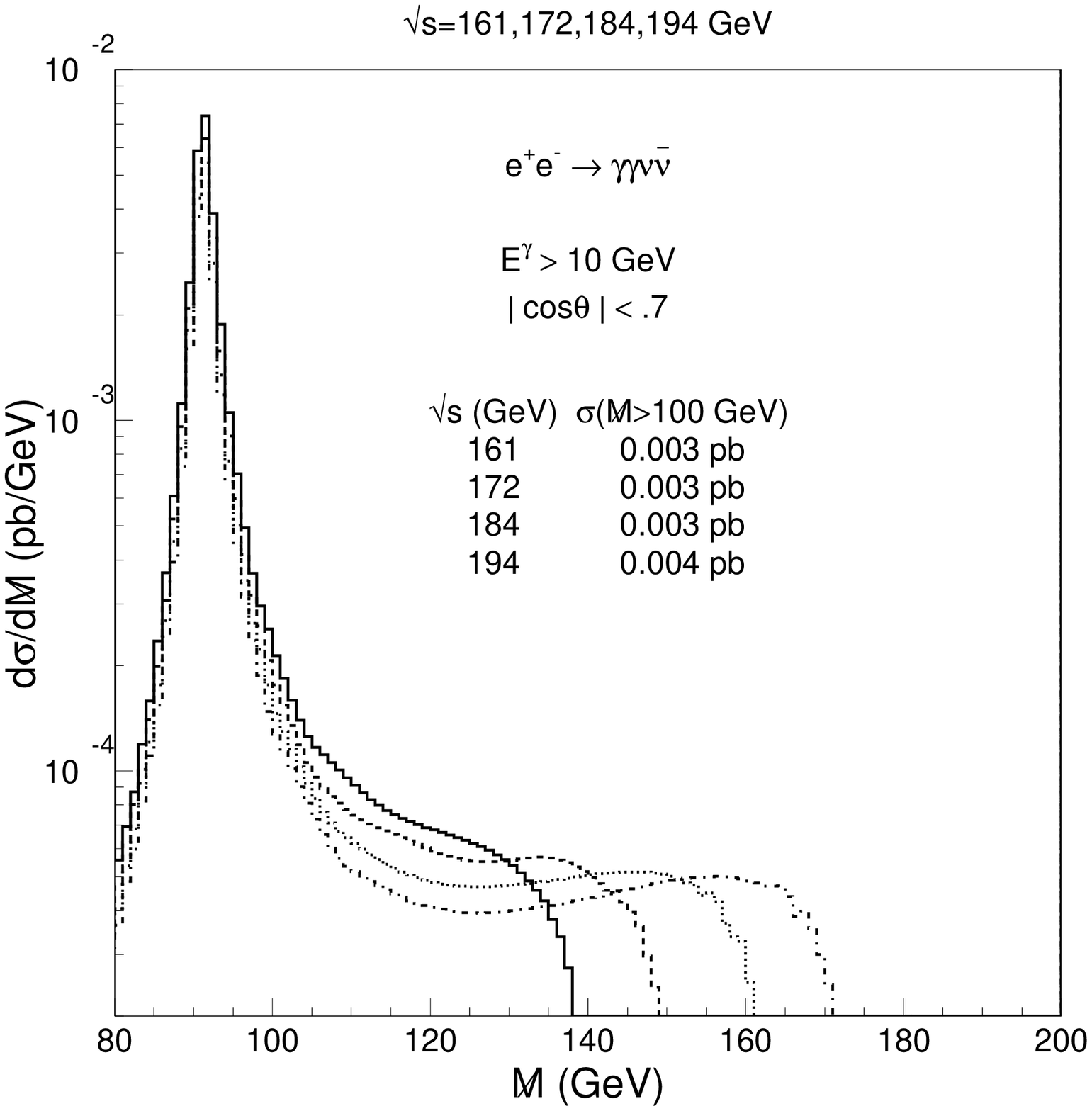}
\vspace*{0in}
\caption{The Missing Mass distribution in pb per GeV bin
for a minimum photon energy of 10 GeV and $|\cos\theta|<.7$ at
$\sqrt{s}$=161, 172, 184, and 194 GeV.  The integrated
cross section above 100 GeV is also shown.}
\label{fig_three}
\end{figure}
\begin{figure}
\centering
\hspace*{0in}
\epsfxsize=5.0in
\epsffile{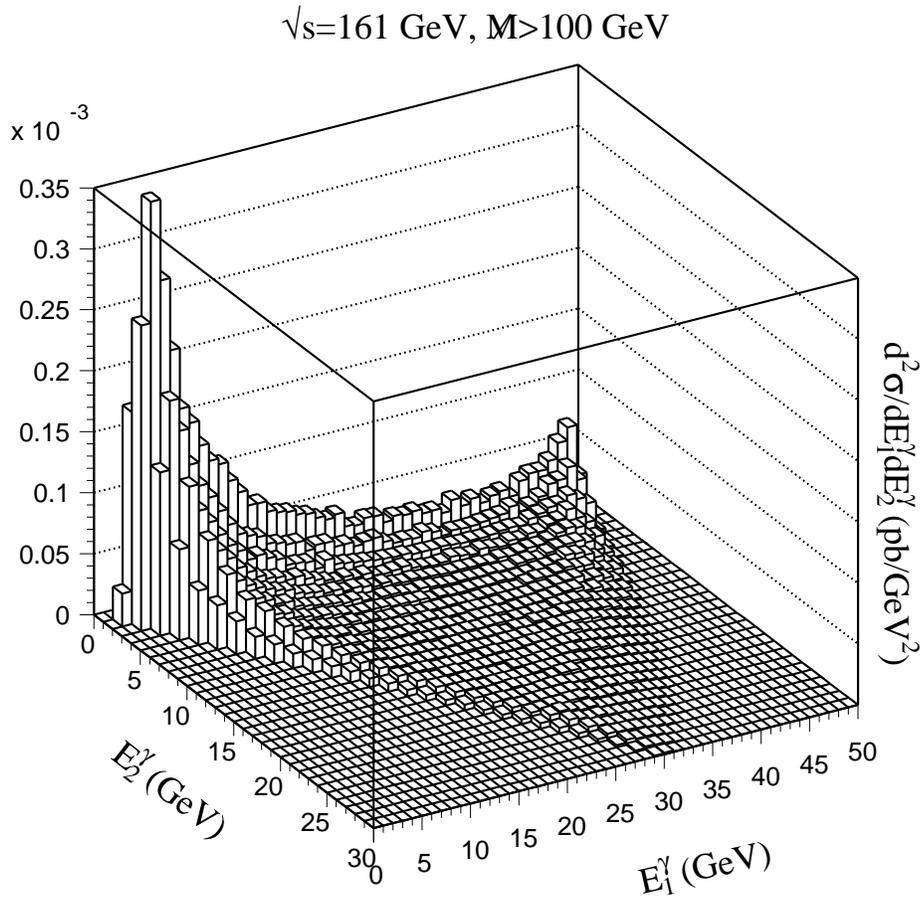}
\vspace*{0in}
\caption{The correlation between the two photon energies
in pb per GeV$^2$ bin at $\sqrt{s}$=161 GeV when
$\slashchar{M}>$ 100 GeV.  Both photons must satisfy
$|\cos\theta|<.7$.}
\label{fig_four}
\end{figure}
\begin{figure}
\centering
\hspace*{0in}
\epsfxsize=5.0in
\epsffile{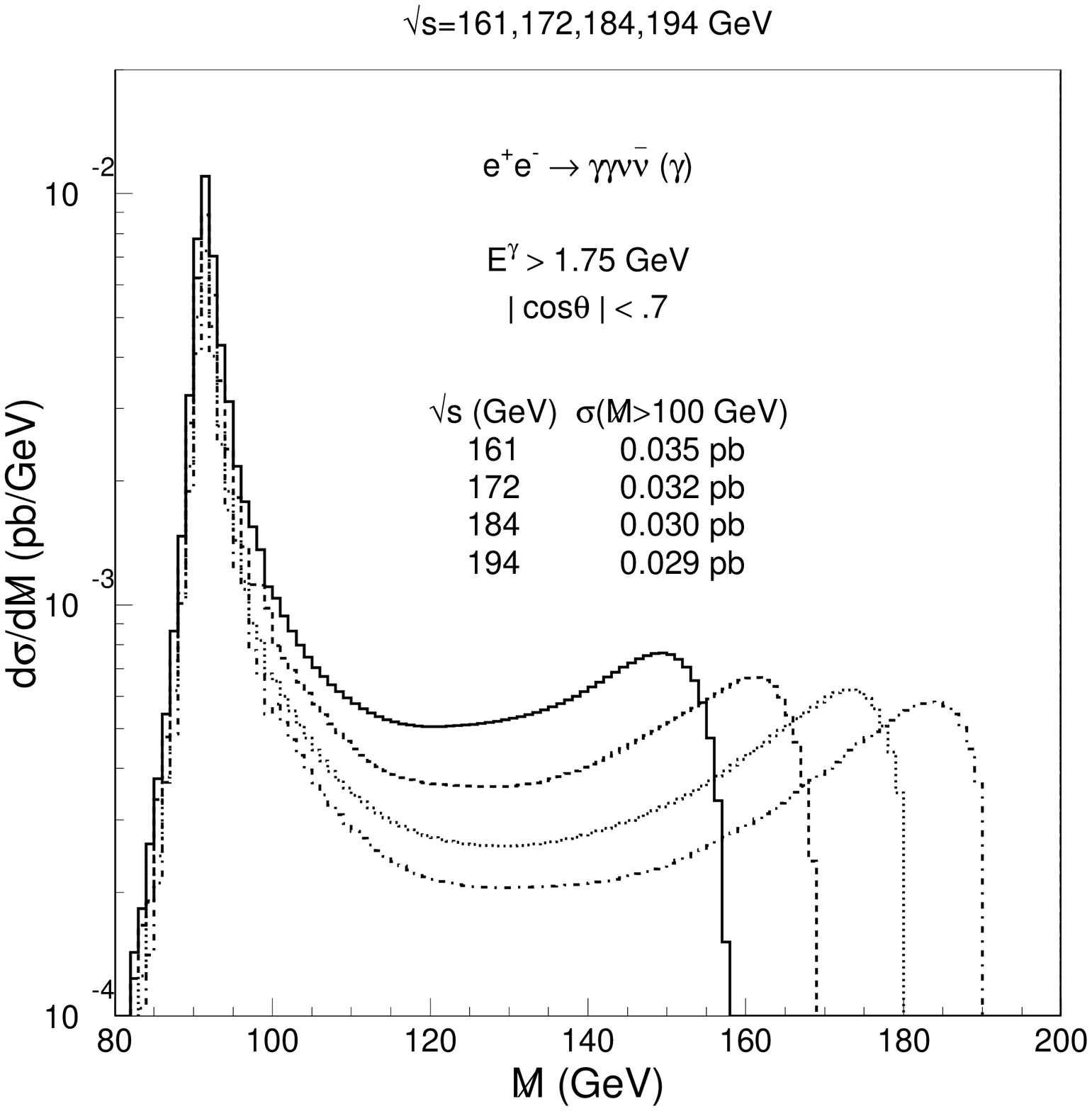}
\vspace*{0in}
\caption{The Missing Mass distribution in pb per GeV bin
for a minimum photon energy of 1.75 GeV and $|\cos\theta|<.7$ at
$\sqrt{s}$=161, 172, 184, and 194 GeV.  The integrated
cross section above 100 GeV is also shown.  
The effect of undetected initial state radiation has been included.}
\label{fig_ones}
\end{figure}
\begin{figure}
\centering
\hspace*{0in}
\epsfxsize=5.0in
\epsffile{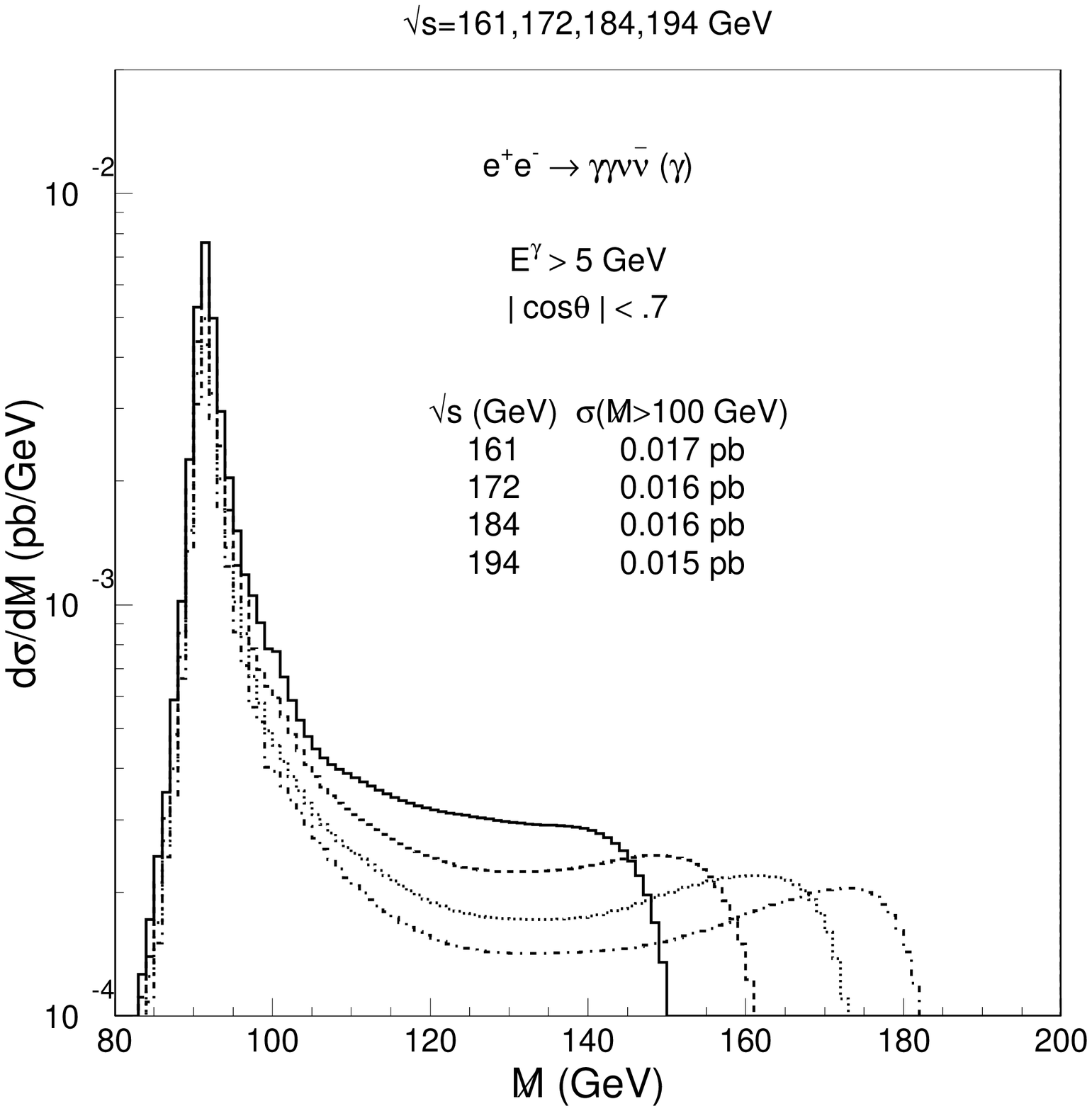}
\vspace*{0in}
\caption{The Missing Mass distribution in pb per GeV bin
for a minimum photon energy of 5 GeV and $|\cos\theta|<.7$ at
$\sqrt{s}$=161, 172, 184, and 194 GeV.  The integrated
cross section above 100 GeV is also shown.
The effect of undetected initial state radiation has been included.}
\label{fig_twos}
\end{figure}
\begin{figure}
\centering
\hspace*{0in}
\epsfxsize=5.0in
\epsffile{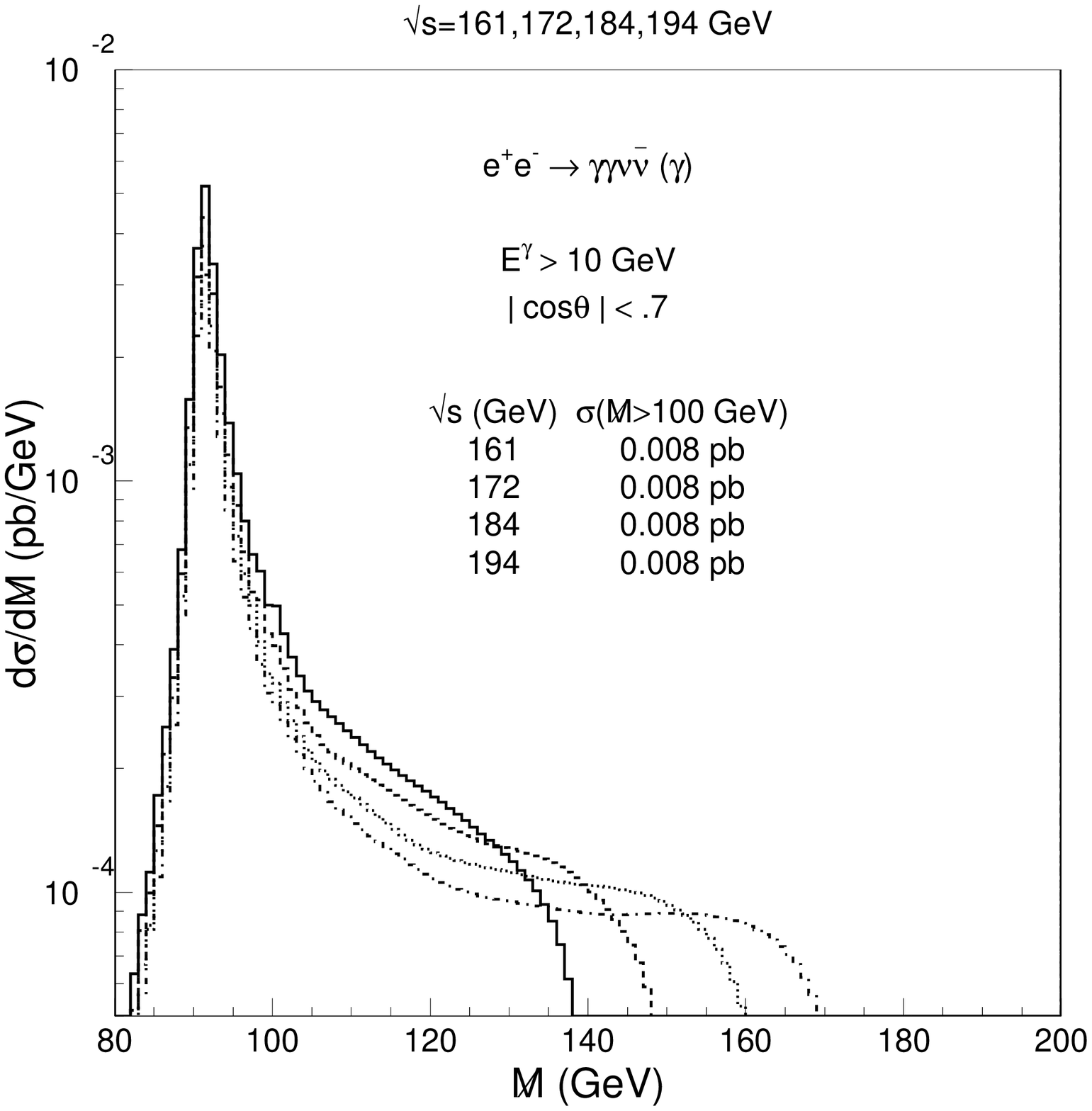}
\vspace*{0in}
\caption{The Missing Mass distribution in pb per GeV bin
for a minimum photon energy of 10 GeV and $|\cos\theta|<.7$ at
$\sqrt{s}$=161, 172, 184, and 194 GeV.  The integrated
cross section above 100 GeV is also shown.
The effect of undetected initial state radiation has been included.}
\label{fig_threes}
\end{figure}
\begin{figure}
\centering
\hspace*{0in}
\epsfxsize=5.0in
\epsffile{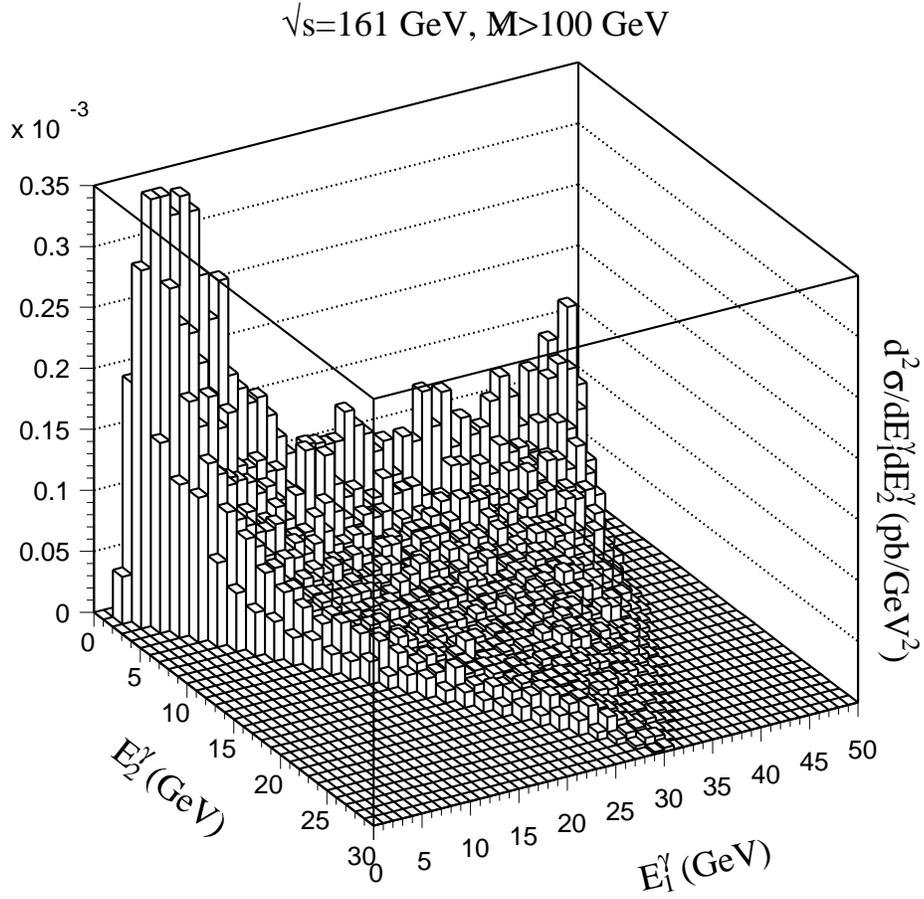}
\vspace*{0in}
\caption{The correlation between the two photon energies
in pb per GeV$^2$ bin at $\sqrt{s}$=161 GeV when
$\slashchar{M}>$ 100 GeV.  Both photons must satisfy
$|\cos\theta|<.7$.
The effect of undetected initial state radiation has been included.}
\label{fig_fours}
\end{figure}
\end{document}